\RequirePackage{lineno}
\documentclass{emulateapj}
\usepackage{natbib}
\usepackage{subfigure}
\usepackage{appendix}
\usepackage[normalem]{ulem}
\begin{document}

\setpagewiselinenumbers
\modulolinenumbers[5]

\title{\large{Gravitational Wave Hotspots:}
 
\small{Ranking Potential Locations of Single-Source Gravitational Wave Emission}}

\author{Joseph Simon$^{1,2}$, Abigail Polin$^{1,3}$, Andrea Lommen$^1$, Ben Stappers$^4$, Lee Samuel Finn$^5$, \linebreak F. A. Jenet$^6$ and B Christy$^1$}
\affil{$^1$Department of Physics \& Astronomy, Franklin and Marshall College, $^2$Physics Department, University of Wisconsin Milwaukee, $^3$Physics Department, University of California at Berkeley, $^4$Jodrell Centre for Astrophysics, University of Manchester, $^5$Department of Physics, The Pennsylvania State University, $^6$Center for Advanced Radio Astronomy, University of Texas at Brownsville}

\begin{abstract}

The steadily improving sensitivity of pulsar timing arrays (PTAs) suggests that gravitational waves (GWs) from supermassive black hole binary (SMBHB) systems in the nearby universe will be detectable sometime during the next decade. Currently, PTAs assume an equal probability of detection from every sky position, but as evidence grows for a non-isotropic distribution of sources, is there a most likely sky position for a detectable single source of GWs?
In this paper, a collection of galactic catalogs is used to calculate various metrics related to the detectability of  a single GW source resolvable above a GW background, assuming that every galaxy has the same probability of containing a SMBHB.
Our analyses of these data reveal small probabilities that one of these sources is currently in the PTA band, but as sensitivity is improved regions of consistent probability density are found in predictable locations, specifically around local galaxy clusters.

\end{abstract}

\section{Introduction}
\setcounter{footnote}{0}
\setcounter{figure}{0}

Pulsar Timing Arrays (PTAs) are collections of millisecond pulsars \citep{Backer90} whose joint timings will show correlations that are a specific signature of gravitational waves (GWs) passing between the Earth and the pulsars \citep{Sazhin78, Detweiler79}. Currently, there are three PTAs working collaboratively to detect GWs in this way \citep{Hobbs10}. These arrays will soon have the sensitivity to detect single extragalactic sources of GWs, \citep{Yardley10, Ellis12} which are resolvable above a GW background \citep{Sesana13CQG}. To increase the efficiency of PTA observations and potentially decrease the time until a detection is made, various groups have considered the optimization of PTA observations \citep{Lee08, Burt10, Lee12}, but all have assumed an equal probability of detection across the sky. While all directions are equally likely to contain a GW source, there should exist more probable locations for the brightest GW source. And while the probability is quite small for there to exist a source that stands out above the background, the analysis in this paper identifies the potential locations of that source using what is currently known about the distribution of galaxies in the local universe, and allows others to enhance discussions on the optimization of a PTA \citep{Anella13}.

Supermassive black hole binary (SMBHB) systems with periods of months to years are thought to be the most important source of gravitational waves \citep{Jaffe03}. Binaries like these form when galaxies containing nuclear black holes merge \citep{Begelman80, Volonteri03}. Corresponsingly, the number of such binaries should be greatest where galaxy mergers are more frequent; i.e., in galaxy clusters. All things being equal, the nearest clusters will play host to the brightest sources. This sugests that, as PTA sensitivities increase, we look toward the largest or richest nearby clusters (Virgo, Fornax, Norma, Perseus and Coma) as the most likely location of the earliest detectable SMBHB source. Here we make this expectation quantitative.

Below in \S 2, we use a compilation of several galaxy surveys to identify the mass, distance, and location of all galaxies within $150~$Mpc.  From this data we estimate the SMBHB mass and the lifetime of GW emission in a detectable PTA band for each galaxy in \S 3, and together with the distance, we identify the probability of the existence of a detectable source in a given direction. \S 4 contains a full explanation of the results with sky position maps. We investigate the probability of detection given an increasing PTA sensitivity in \S 5 and a summary of our findings is found in \S 6.

\section{Extragalactic Data Base}

We require knowledge of the mass and distance of each galaxy in order to estimate the amplitude of the GWs emitted by a SMBHB that may exist at the center of that galaxy. We start by searching the Extragalactic Distance Database (EDD), created by \citet{Tully09}.  This database is a compilation of many extragalactic surveys with the intention of compiling all visible galaxies within $140/h$~Mpc (z=0.03), where $H_o = 72~h$~km/s/Mpc. The EDD\footnote{\label{edd}http://edd.ifa.hawaii.edu} recently updated its records with the 2M++ galaxy reshift catalog which reaches $90\%$ complete out to 200$/h$ Mpc and is conservatively complete to 60$/h$ Mpc \citep{Lavaux11}. This gives our sample the same completeness. Additionally, the EDD was updated with an extensive survey of all galaxies within 11 Mpc \citep{Karachentsev13}, later in the paper local sources will be highlighted and this recent addition to the EDD gives us great confidence in our ability to talk about neighboring galaxies.

The Lyon-Meudon Extragalactic Database\footnote{http://leda.univ-lyon1.fr} (LEDA) \citep{Paturel03} is the largest database that the EDD draws on, but while LEDA compiles over three million objects \citep{Vauglin06} the EDD only gathers information from LEDA for galaxies that are found in other surveys \citep{Tully09} which at the time of this paper was just under a hundred thousand galaxies$^{\ref{edd}}$.  Both the EDD and LEDA take advantage of the labeling mechanism started in the Principal Catalogue of Galaxies (PGC) \citep{Paturel89}.  The PGC number of every known galaxy is used to access the same galaxy across several surveys and to prevent any single galaxy from being counted more than once. 

To obtain an accurate distance, the databases use various methods based on the known parameters of each galaxy.  There are two main distance moduli calculated in LEDA, ``mod0", which is calculated from a distance catalog using the Tully-Fisher relation or the Faber-Jackson relation, and ``modz", which is calculated using redshift \citep{Paturel97}. The EDD primarily uses the Tully-Fisher relation to get distances, but compares the results with other distances to assure a common scale \citep{Tully09}. Distance measurement using the Tully-Fisher relation is only available for about $6\%$ of galaxies, with the rest being derived from redshift measurements.

\subsection{Estimating Black Hole Masses}

\begin{figure}
	\centering
	\includegraphics[width=0.45\textwidth]{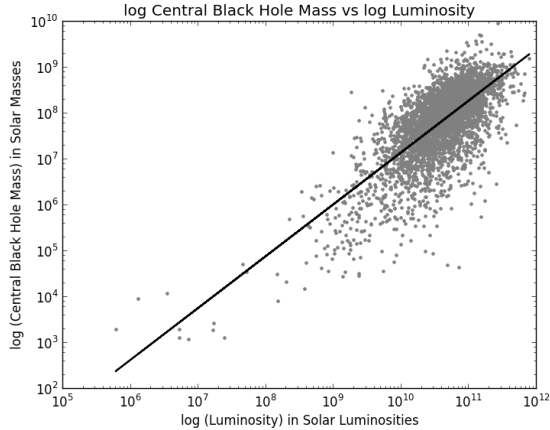}
	\caption{A strong correlation is shown between the mass of a galaxy's central black hole and that galaxy's luminosity, specifically in the high black hole mass range, $>10^7$ solar masses. The above log-log plot shows a clear trend that is best represented by the line $y=(1.13 \pm 0.02)x - 4.17$. This plot was made using all galaxies with a known $\sigma$ and a well defined B-band luminosity found in the extragalactic databases.}
	\label{fig:bhvl}
\end{figure}

We calculate the total central black hole mass of a galaxy using the M-$\sigma$ relation when an accurate central velocity dispersion ($\sigma$) is found. This is available for $5\%$ of the galaxies. In an attempt to expand the pool of usable galaxies, B-band luminosity is used as a surrogate for mass.  In LEDA, most known galaxies have a calculated total B-band luminosity, which we converted to a total luminosity using a bolometric correction \citep{Buzzoni05}. The number of galaxies for this work increased from 4,258 to 83,816. Fig \ref{fig:bhvl} compares the luminosity and black hole mass for galaxies with known $\sigma$. While the relationship is most likely non-linear in lower mass regions, since PTAs are only sensitive to a certain range of black hole masses, $>10^{7}$ solar masses \citep{Sesana10}, we can ignore the low mass trend and only use the clear linear trend in the PTA sensitive region. We advise the reader that the line looks like a poor fit  to the data because of the invisible density gradient in the gray region of the plot.

For consistency, all gathered luminosities were converted to black hole mass using the above trend line which gives the power law,
\begin{eqnarray}
	\label{eqn:bhvl}
	M_{\rm BH} = 10^{-4.17} {\rm L}^{1.13},
\end{eqnarray}
where L is the corrected luminosity of each galaxy.

\subsection{Sample Size}

\begin{figure}
	\centering
	\includegraphics[width=0.45\textwidth]{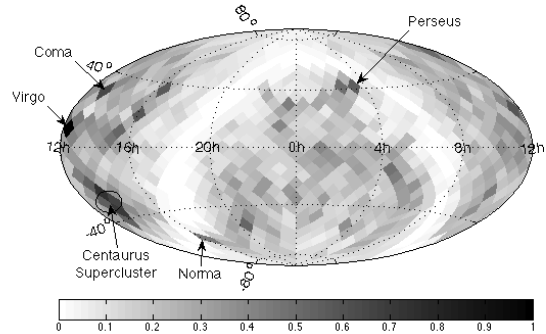}
	\caption{The number density of total galaxies found with a recorded B-band luminosity are binned in equal-area pixels in order to detect any bias in the data.  Empty areas of the plot are caused by the zone of avoidance, and dark pixels contain galaxy clusters.  The distribution across the sky shows no noticeable survey bias.}
	\label{fig:totalgalaxies}
\end{figure}

\begin{figure}
	\centering
	\includegraphics[width=0.45\textwidth]{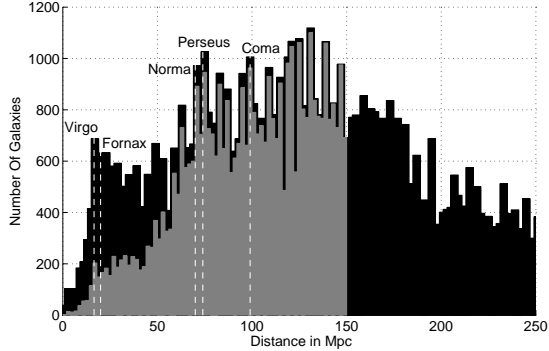}
	\caption{The number of galaxies as a function of distance is shown above. The total number grows with distance as expected out to 150 Mpc, with spikes at noted galaxy clusters. The gray region shows the number of galaxies after we require galaxies have a central black hole mass larger than $10^7$ solar masses, and have distances less than 150 Mpc out to which the surveys are roughly complete.}
	\label{fig:disthist}
\end{figure}

In Fig~\ref{fig:totalgalaxies}, we plot all found galaxies with a recorded B-band luminosity to look for survey biases. This plot was created with equal-area pixels generated using MEALpix\footnote{\label{Mpix}Program developed by GWAstro Research Group and available from http://gwastro.org}. The empty areas in the plot are caused by the Milky Way galaxy, whose plane renders these areas of the sky unobservable, also called the zone of avoidance. The largest value pixels in this plot, appearing as black and dark gray, contain galaxy clusters. Significant clusters are labeled on the plots throughout this paper. The Virgo cluster contains approximately 1300 galaxies \citep{Binggeli85}; the Fornax cluster contains around 60 galaxies \citep{Jordan07}; the Norma cluser contains around 600 galaxies \citep{Woudt08}; the Perseus cluster contains about 500 galaxies \citep{Brunzendorf99}; and the Coma cluster contains more than a thousand galaxies \citep{Hammer10}. Since all pixels not obstructed by the plane of the Milky Way contain a number of galaxies that are within the same order of magnitude, we deem the distribution across the sky to be reasonably non-bias, particularly with respect to nearby sources, which are most important to us. Fig~\ref{fig:disthist} contains a plot of all galaxies with a recorded B-band luminosity in the combined databases as a histogram over distance. We label the mean distance of some galaxy clusters, and in general the larger spikes in galaxies are due to clusters.  The number of galaxies grow out to approximately 150 Mpc, and then fall off inversely with distance. From the completeness of the EDD, we feel confident that in this paper we achieve a reasonably accurate representation of our local universe, i.e. the galaxies within 150 Mpc.

\begin{figure}
	\centering
	\includegraphics[width=0.45\textwidth]{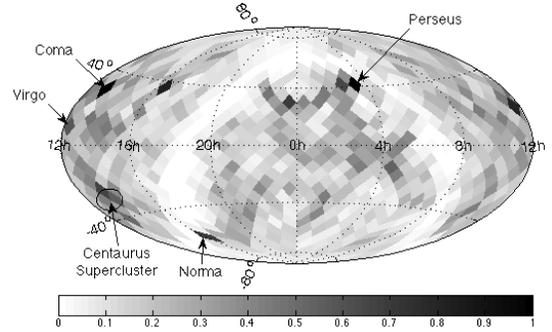}
	\caption{This is the distribution of galaxies that were found from the extragalactic databases to have a total central black hole mass larger than $10^7$ solar masses and are within 150 Mpc. In comparison to Fig~\ref{fig:totalgalaxies}, this plot clearly shows the distribution of number density of galaxies in the local universe to be dominated by galaxy clusters. This distribution shows no clear bias and is the distribution used in the rest of the paper.}
	\label{fig:massdistcutgalaxies}
\end{figure}

Now that we have a significant sample size of galaxies, we cut down to only the galaxies with luminosities great enough to potentially harbor a detectable source; galaxies with central black holes larger than $10^7$. In an attempt to remove bias to our closest neighbors, we remove the Andromeda galaxy and our galaxy from the dataset as well. The effect of this mass cut can be seen on the gray histogram overlayed on Fig~\ref{fig:disthist}, which leaves us with 75,486 galaxies. The region of sky within 50 Mpc, is the most effected by this cut, which makes sense given that less luminous galaxies will only be observable at smaller distances. The Virgo cluster is the only galaxy cluster strongly effected by this cut for similar reasons. After the mass cut was made, we cut down to only the galaxies within 150 Mpc of the Milky Way, which leaves us with 40,560 galaxies, just under half of the starting number. The effect of this distance cut verses sky position can be seen in Fig~\ref{fig:massdistcutgalaxies}.

\section{Analysis}

To estimate the detectibility of GWs with PTAs, we use the SMBHB mass and the distance of each galaxy to calculate the GW signal strength from each potential source. We also calculate the lifetime of emission in a detectable PTA band, and estimate the number of detectable sources. This allows us to identify the probability of the existence of a source at a given amplitude in a given direction.

We start with the simplified assumption that all galaxies contain a binary with equal mass black holes. Since we are only trying to estimate the ``hotness" of sky positions relative to each other, we work in proportionalities. The overall factors are irrelevant at this stage of our understanding. A more mathematical description of these proportionalities can be found in the appendix.

We use MEALpix$^{\ref{Mpix}}$ to divide the sky into equal-area pixels, and associate each galaxy with a particular pixel based on its sky location.

If we assume a given galaxy contains or contained a SMBHB then the probability that the SMBHB exists now is the ratio of the binary's lifetime, $\tau$, to the age of the universe.  If we further assume that the source is detectable as long as the SMBHB exists then the expectation value of the number density of sources in a certain pixel, $\langle N_7 \rangle$, is directly proportional to the lifetime, $\tau$, of all sources in that pixel;
\begin{eqnarray}
	\langle N_7 \rangle \propto \sum_{i}^{\rm{N}} \tau_i ,
\end{eqnarray}
where N is the total number of galaxies in a given pixel and $i$ represents a particular galaxy in that pixel. $\langle N_7 \rangle$ serves as one of the metrics we use to characterize the "hotness" of the GW sky. However, $\langle N_7 \rangle$ says nothing about the relative strength of sources, just the number of them.  We therefore also use the metric $\langle P \rangle$ that is proportional to the expectation value of the power in GWs emitted from a particular pixel.  Consider the GW power contained in the residual response from a single source, P; the response in pulsar timing to a gravitational wave of strain $h$ is proportional to $h$ times the period of the binary \citep{Jenet04}.  However, we assume that all the SMBHBs have the same period, and drop the length of the period from our calculations.  Therefore, the residual response is proportional to $h$.
The power in the pulsar timing residuals is the square of this response:
\begin{eqnarray}
	{\rm{P}} \propto h^2 .
\end{eqnarray}
The total energy emitted over the lifetime of the source is proportional to the power times the lifetime of the source, $\tau$. Therefore
\begin{eqnarray}
	\int {\rm{P}} \,dt \propto h^2 \tau
\end{eqnarray}
where the integral is over all time. When we add up all the sources in a particular pixel we get
\begin{eqnarray}
	\int P_{\rm{pixel}} \,dt \propto \sum_i^{\rm{N}} h_i^2 \tau_i.
\end{eqnarray}
where N is the total number of galaxies in a given pixel and $i$ represents a particular galaxy in that pixel.  In essence, this is the total amount of energy a PTA can expect to receive from this pixel over all time, and after dividing by the age of the universe would be the average power in the pixel. Therefore $\int P_{\rm{pixel}} \,dt$ is also proportional to the expectation value of the power from this pixel and we have an expression for our second metric, $\langle P \rangle$,
\begin{eqnarray}
	\langle P \rangle \propto \sum_i^{{\rm{N}}} h_i^2 \tau_i.
\end{eqnarray}

The above equations only depend on the GW strain, $h$, and the lifetime of each SMBHB, $\tau$.  We use the standard dipole approximation from \citet{Jenet04} to
estimate the strain $h$;
\begin{eqnarray}
	\label{eqn:strain}
 	h \propto \frac{M_c^{5/3}}{d},
\end{eqnarray}
where $M_c$ is the chirp mass of the SMBHB. The lifetime of the source, $\tau$, is given by:
\begin{eqnarray}
	\label{eqn:lifetime}
	\tau \propto \frac{1}{M_c^{5/3}}.
\end{eqnarray}

The chirp mass of a binary system, $M_c$, is proportional to the total mass of the binary, $M_{\rm T}$, since we have already assumed that the masses in the binary are roughly equal, 
\begin{eqnarray}
	M_c = M_{\rm T} (\frac{m_1 m_2}{{M_{\rm{T}}}^2})^{5/3} = 0.4 M_{\rm T}.
\end{eqnarray}

We can now rewrite both the number of detectable sources and the GW signal strength in terms of $M_{\rm T}$ and $d$, which are the observed quantities gathered in \S 2.
\begin{eqnarray}
	\label{eqn:dect}
	\langle N_7 \rangle \propto \sum_{i}^{\rm{n}} \frac{1}{M_{\rm T}^{5/3}} \\
	\label{eqn:power}
	\langle P \rangle \propto \sum_{i}^{\rm{N}} \frac {M_{\rm T}^{5/3}}{d^2}
\end{eqnarray}
Maps of the above quantities can be found in the next section.

\begin{figure}[h]
	\centering
	\includegraphics[width=0.45\textwidth]{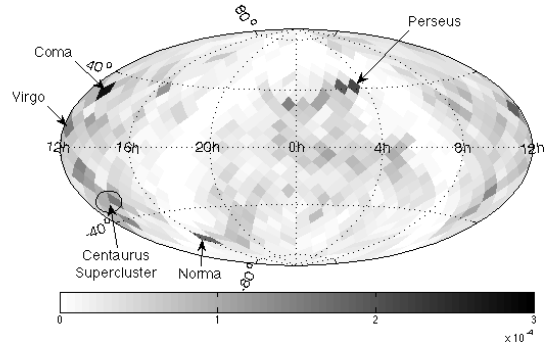}
	\caption{The probability of a detectable source currently being in the PTA band is plotted in each pixel on this plot. The darker regions of the map show where there is a larger density of detectable sources. The darkest pixels in this plot, which correspond to the largest probability, are the pixels containing the Coma and the Perseus cluster.  These regions of the sky are understood as the directions with a greater number density of sources containing a GW source when considering longer observations.}
	\label{fig:f1}
\end{figure}

\begin{figure}[h]
	\centering
	\includegraphics[width=0.45\textwidth]{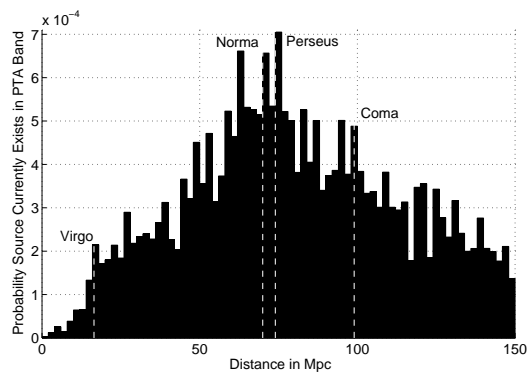}
	\caption{The probability of a detectable source currently being in the PTA band verses distance is plotted above to give a sense of spatial depth to the sky plot. As in the sky plot, the Perseus cluster appears as a region with a larger probability of detectable sources. The Coma cluster still appears as a significant region along with the Norma cluster.}
	\label{fig:f1hist}
\end{figure}

\section{Results}

There are many factors that contribute to the detection of a GW signal by a PTA, and so measuring only one quantity is insufficient to assess the likelihood of detecting a GW source in a particular region of the sky.  For example, when calculating the number of detectable sources we rank the galaxies in terms of their likelihood of being `on' during an observation, while when calculating GW signal strength we rank the sources in terms of the overall power they are expected to contribute to the pulsar timing band over the course of their lifetimes. These plots give different pictures of our local universe and are both needed to accurately understand a PTAs probability of making a detection.

\subsection{Number Density of Detectable Sources}

One way of finding probable locations for PTA detection is to look at the expectation value of the number of sources in any given direction $\langle N_7 \rangle$, which we estimate using Eqn~\ref{eqn:dect}. 
This value is proportional to the probability of a detectable source currently being in the PTA band in a given pixel. Using the equations in the appendix, the amount of time each potential source spends in the PTA band is calculated, which is converted to a probability by dividing by the age of the universe. We find the total probability of a single source currently being detectable to be $0.023$, with the `brightest' location on the sky having a $2.9 \times 10^{-4}$ probability of currently containing a single source that stands out about the background in the PTA band.
The probability of a detectable source currently being in the PTA band is plotted verses sky position in Fig~\ref{fig:f1}, and as a function of distance in Fig~\ref{fig:f1hist}. These plots are dominated by galaxy clusters in the distance range of 50 to 100 Mpc. Specifically the Coma, Norma and Perseus clusters, which all contain numerous massive galaxies, while also being at a close enough distance for those galaxies to be resolvable by a PTA.
These plots highlight a region of space between 50 and 75 Mpc where a large portion of galaxies are a part of the Centaurus supercluster. This region of space which stretches down towards the Norma cluster, is partially in the zone of avoidance and contains the Great Attractor \citep{Kocevski07}. While optical observations will likely not reveal the Great Attractor, PTAs have the potential to discover the source of this attraction from GW observations.

\begin{figure}[h]
	\centering
	\includegraphics[width=0.45\textwidth]{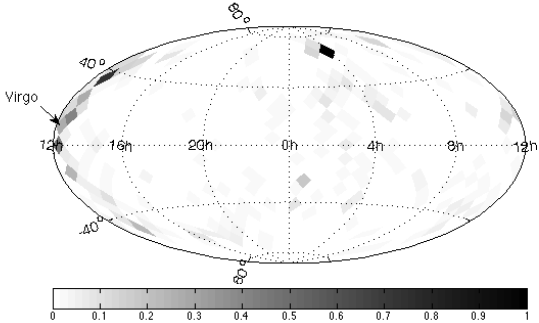}
	\caption{The expectation value of GW Power, $\langle P \rangle$, from Eqn~\ref{eqn:power} is plotted verses sky position. This plot is dominated by a handful of local sources, including the Virgo cluster.}
	\label{fig:f2}
\end{figure}

\begin{figure}[h]
	\centering
	\includegraphics[width=0.45\textwidth]{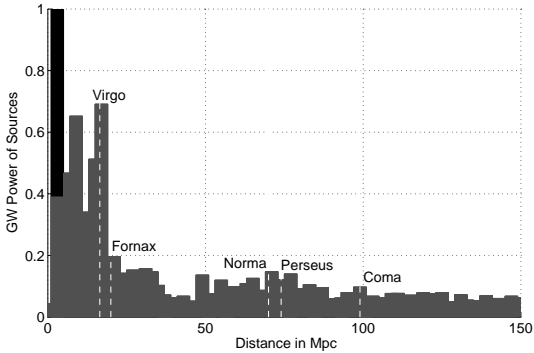}
	\caption{The above plot of $\langle P \rangle$ calculated using Eqn~\ref{eqn:power} verses distance gives a sense of spatial depth to the sky plot in Fig~\ref{fig:f2}. This plot shows that a local source has the potential to dominate a PTAs search of single GW sources. The black peak is the brightest source of $\langle P \rangle$, and we removed it to make the gray overlay showing the importance of local powerful sources.}
	\label{fig:f2hist}
\end{figure}

\subsection{GW Power}

$\langle P \rangle$ is estimated using the total GW power for each source in a particular pixel and at a particular distance, integrated over its lifetime, which effectively weights the power from each source with the probability of whether or not it will be caught `on' during an observation. Estimated using Eqn~\ref{eqn:power}, $\langle P \rangle$ is plotted verses sky position in Fig~\ref{fig:f2} and as a function of distance in Fig~\ref{fig:f2hist}. These plots highlight a handful of local sources that have the potential to dominate a PTAs detection of a single GW source. In an attempt to show just how dominating a single source can be, we removed the brightest source, (RA $= 4$h, dec $= +60^o$, distance $= 2.2$ Mpc) and overlayed the total power as a function of distance in Fig~\ref{fig:f2hist}. With the largest source of GW power removed from the plots, a small number of galaxies, specifically those around the Virgo cluster, continue to dominate the picture. Unlike the $\langle N_7 \rangle$ plots which highlight the region of space between 50 and 100 Mpc, the galaxies with the largest $\langle P \rangle$ values are within a distance of 20 Mpc.

\begin{figure*}
	\centering
	\subfigure[Case A]
	{
		\includegraphics[width=0.3\textwidth]{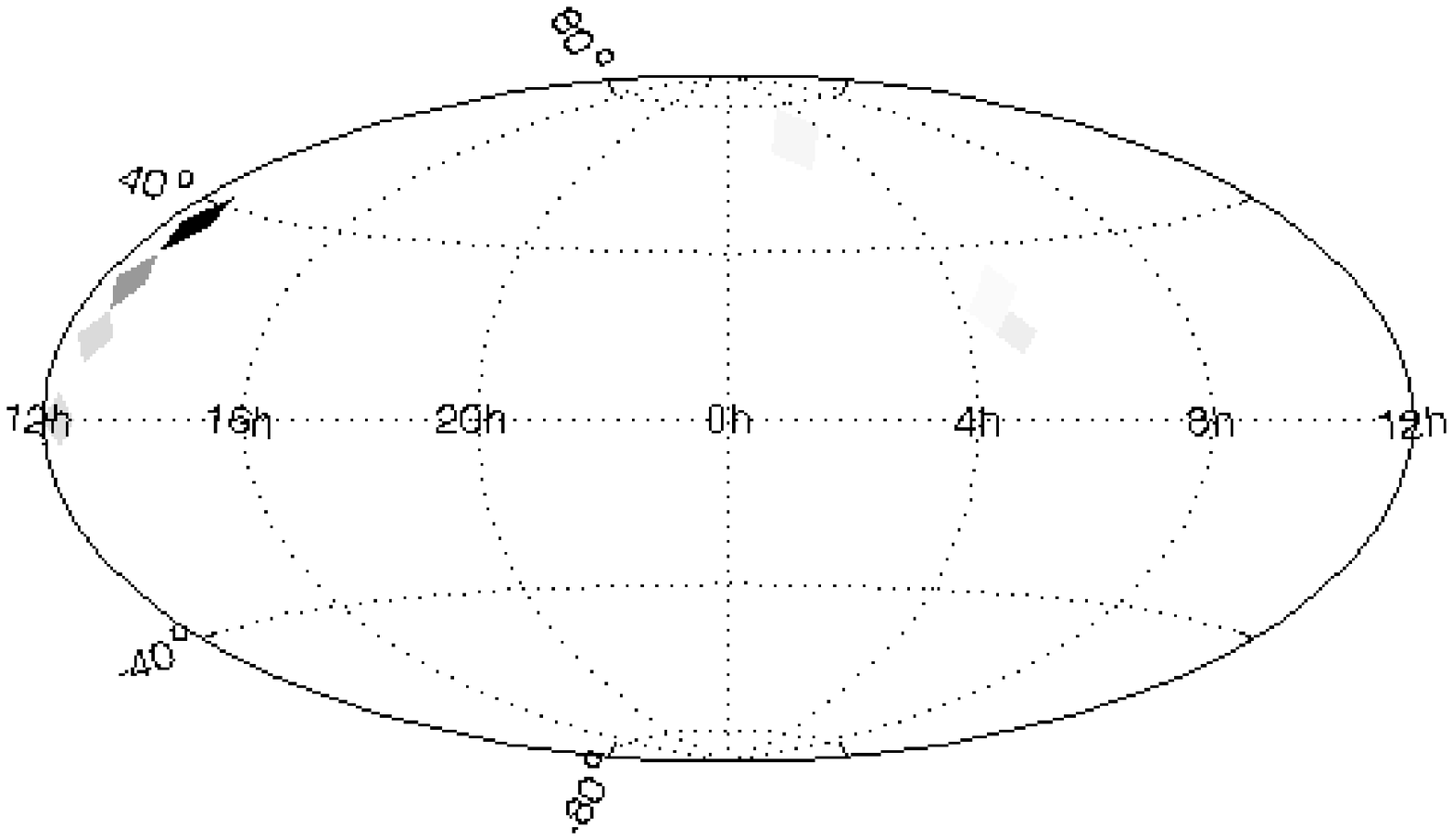}
		\label{fig:ha}
	}
	\subfigure[Case B]
	{
		\includegraphics[width=0.3\textwidth]{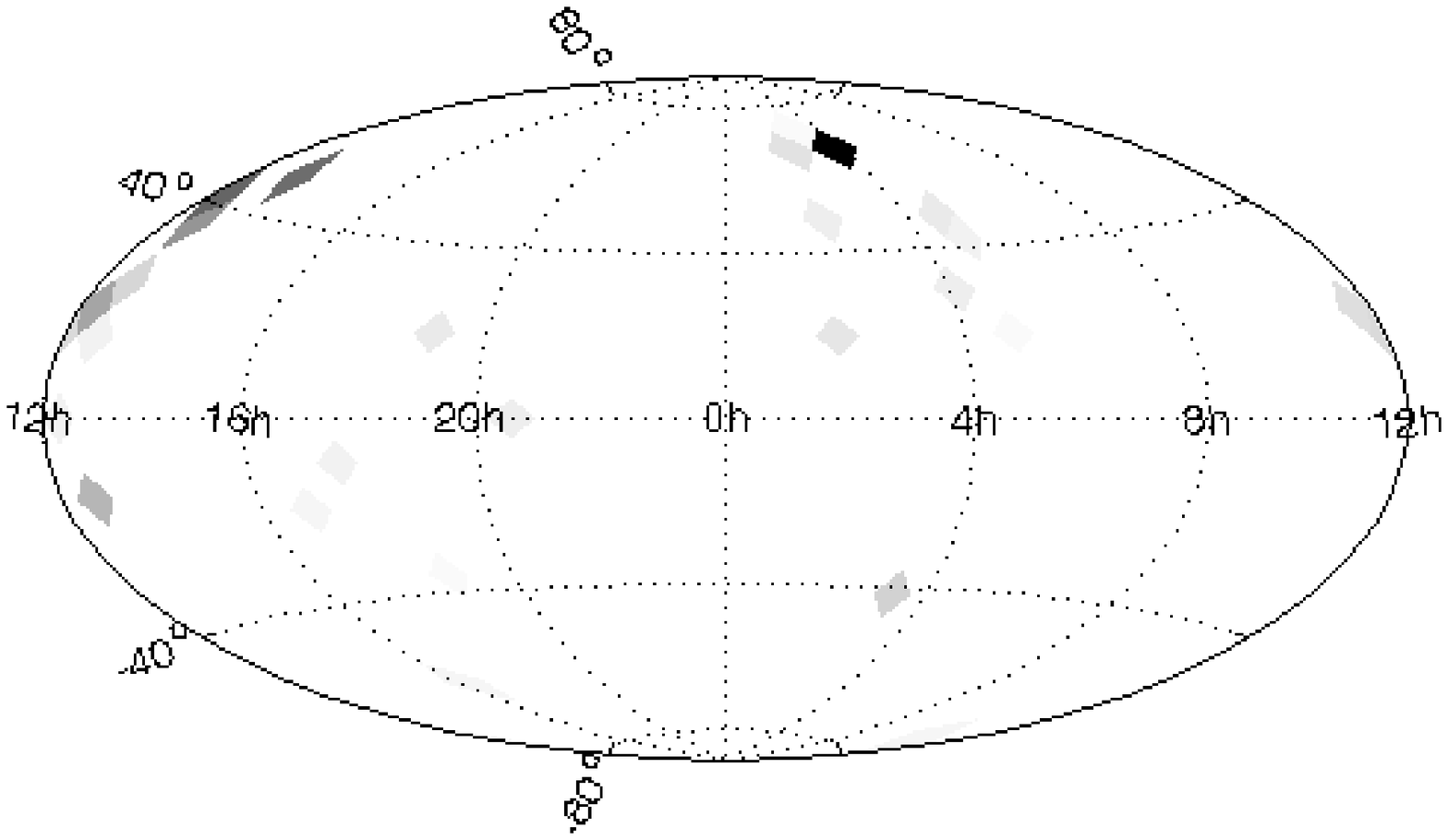}
		\label{fig:hb}
	}
	\subfigure[Case C]
	{
		\includegraphics[width=0.3\textwidth]{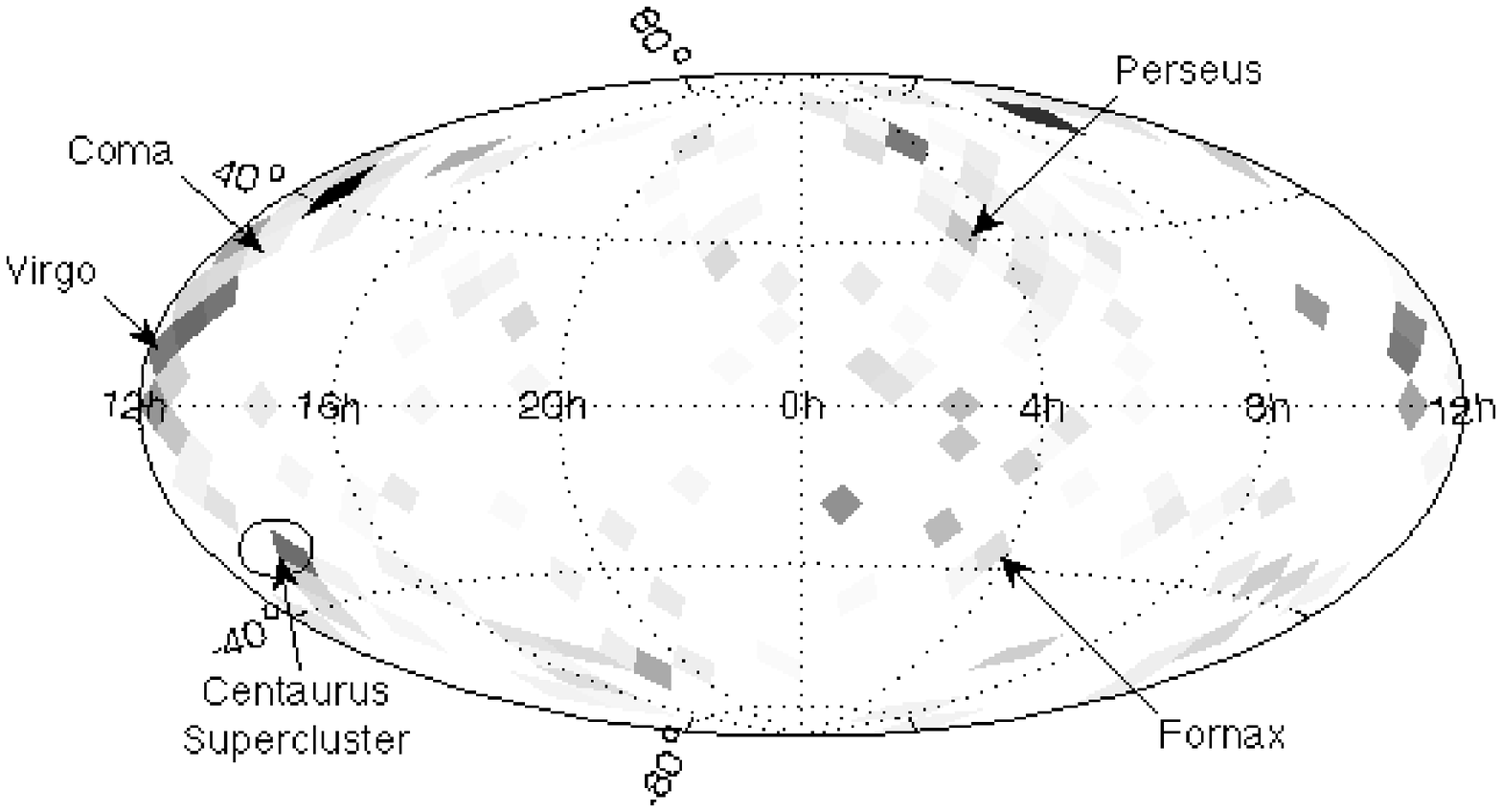}
		\label{fig:hc}
	}\\
	\subfigure[Case A]
	{
		\includegraphics[width=0.3\textwidth]{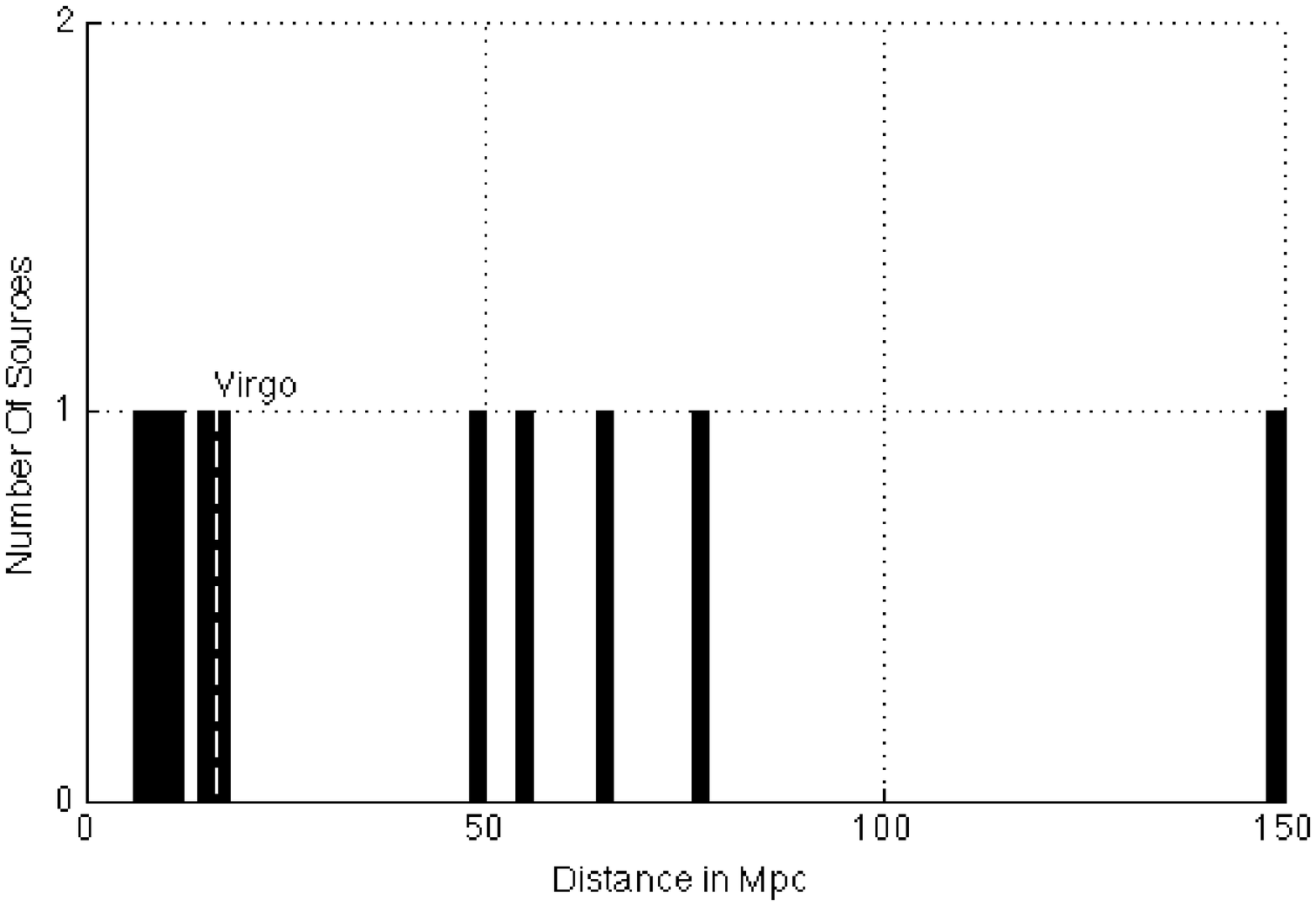}
		\label{fig:hahist}
	}
	\subfigure[Case B]
	{
		\includegraphics[width=0.3\textwidth]{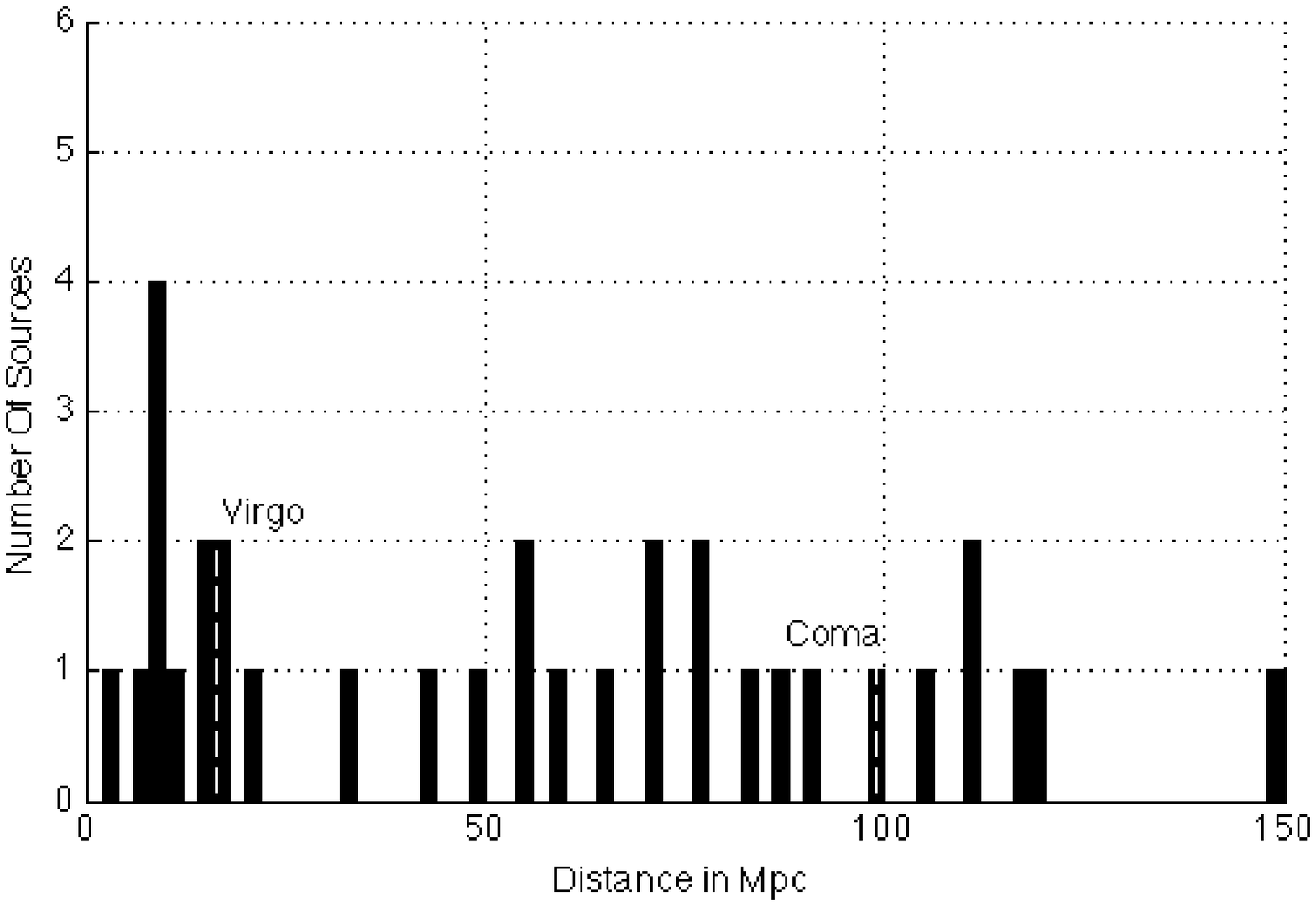}
		\label{fig:hbhist}
	}
	\subfigure[Case C]
	{
		\includegraphics[width=0.3\textwidth]{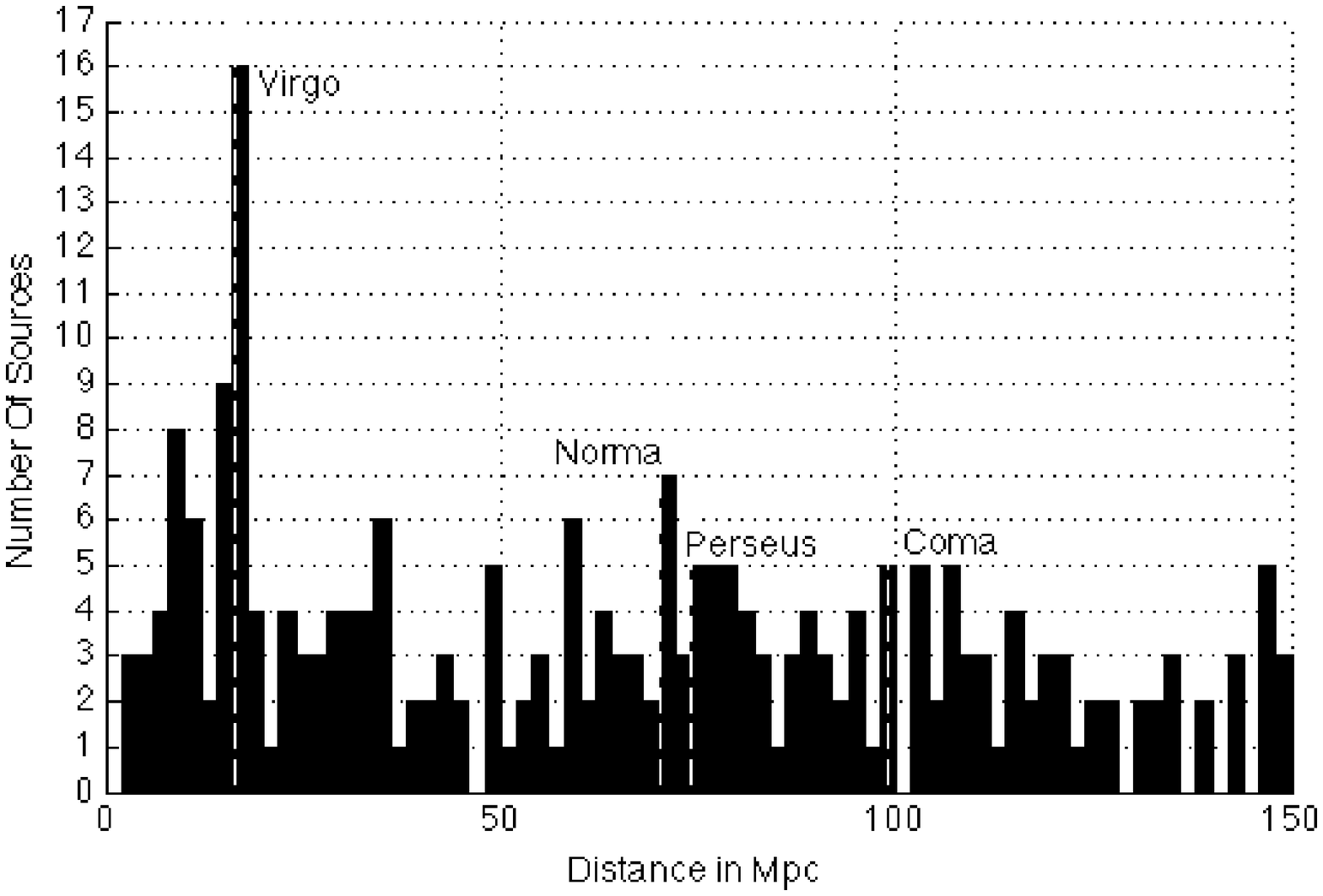}
		\label{fig:hchist}
	}
	\caption{Given a diminishing PTA sensitivity, these are the plots of detectable PTA sources. The signal sensitivity cut off is determined via the GW strain of each galaxy, $h$, calculated from Eqn~\ref{eqn:strain}.  Case A starts with ten potential sources and each subsequent case increases the PTA sensitivity by a factor of $\sqrt{10}$. The probability of detecting any individual source is proportional to that source's lifetime, $\tau$ Eqn~\ref{eqn:lifetime}. This probability is plotted verses sky position, and the number of galaxies at a given distance is plotted for each case as a reference to provide spatial depth. At higher sensitivities, the plots are dominated by a handful of sources whose lifetimes are significant fractions of the total lifetime of all detectable sources, with many of these sources being within 20 Mpc. Once the sensitivity is lowered, the sources become more evenly distributed with a ``hotspot" appearing around the Virgo cluster.}
	\label{fig:hprob}
\end{figure*}

\section{Probability of Detection With A PTA}

For a PTA to detect a source, that source must not only be emitting GW radiation in a detectable band, but also at a sufficient amplitude. While the above plots provide an accurate picture of the local landscape for PTA detectable sources, we have not yet taken into account the threshold of a PTA to detect these sources. 

Any PTA will have a minimum detectability threshold defined by its sensitivity. This threshold is directly related to the strain amplitude, $h$ of a potential GW source.  Over time, the PTAs will continually lower this threshold as improvements to timing, data analysis, and amount of data continue. Eventually, virtually all potential sources in Section 4 will be detectable, and the maps presented are the best guides of where to focus efforts.  However, the first sources detected will be the sources with the largest $h$.

In this section, we make an educated guess as to where those first sources might be located based on our collected data.  We start by considering the ten brightest potential sources, which we label case A.  To mimic improvements to the sensitivity, we also consider two more cases, B $\&$ C, that reduce the initial cutoff value by subsequent factors of $\sqrt{10}$. Our goal is to predict which of these bright sources will be detected first. Therefore, once all the sources above a given threshold have been identified, a probability is assigned to each source by considering its lifetime over the total lifetime of the sample, or $\langle N_7 \rangle$. 

In case A we consider the 10 sources with the highest amplitude $h$ according to Eqn~\ref{eqn:strain}.
The likelihood of one of those sources being detected by a PTA is plotted in Fig~\ref{fig:ha}. The pixel with the most likelihood contains two sources. Each of the two sources in that pixel has a lifetime of about one quarter of the total lifetime of all the detectable sources. In Fig~\ref{fig:hahist}, we see that half of these sources are within 20 Mpc.

When the sensitivity is increased by a factor of $\sqrt{10}$ in case B, there are thirty three potential sources.  Fig~\ref{fig:hb} plots the likelihood that one of these sources is detected. As in case A, the plot is dominated by one very likely pixel, however in this case there is only one source in that pixel and it contains 20$\%$ of the total lifetime of all detectable sources. This source has the largest $\langle P \rangle$ value, discussed in \S 4.2, and while it has a smaller strain than the initial ten, its lifetime is six times longer than any of the potential sources in case A. This source is at a distance of 2.2 Mpc, and the distance distribution of the other sources is plotted in Fig~\ref{fig:hbhist}, where a third of the potential sources are within 20 Mpc.

In case C, the sensitivity is increased one order of magnitude from case A, and there are 237 potential sources. The likelihood of detecting a certain source is plotted in Fig~\ref{fig:hc}. This plot is unlike the Case A and Case B plots since it has sources that are more distributed in clusters. The pixel with the largest total lifetime contains three sources, and contains five percent of the total lifetime of all detectable sources at this cutoff. Fig~\ref{fig:hchist} shows that these sources are evenly distributed throughout distance with a small ``hotspot" appearing around the distance of the Virgo cluster.

Overall, the first two cases reveal very specific sources that dominate the likelihood of detection. It is important to note that as sensitivity is lowered the new sources that will become detectable have the potential for significantly longer lifetimes and these sources are more likely to be found in galaxy clusters, specifically the Virgo cluster.

\section{Summary}

In this paper, data is gathered from a compilation of at least $90\%$ complete galactic surveys out to 200 Mpc. These galaxies are all assumed to contain a SMBHB with equal mass black holes, and the total central mass is calculated using B-band luminosity as a surrogate for mass. This data set is then cut down to only detectable sources (i. e. sources with a total central mass larger than $10^7$ and with in a distance of 150 Mpc). From this data, two metrics are used to estimate the ``hotness" of the GW sky: $\langle N_7 \rangle$, which is proportional to the number density of sources in a given direction; and $\langle P \rangle$, which is proportional to the expectation value of the power in GWs emitted from a particular direction.
Using $\langle N_7 \rangle$, we are able to calculate the specific probability that each pixel contains a currently radiating GW source in the PTA band. While this number is very small for any individual pixel, we find a total probability of $0.023$ that one of the galaxies we considered is detectable. The `brightest' location on the sky has a $2.9 \times 10^{-4}$ probability of currently containing a single source that stands out about the background in the PTA band. Overall the distribution of single sources potentially detectable by a PTA has a larger number density around local galaxy clusters. While the GW signal strength is dominated by a handful of sources, with the region of sky around the Virgo cluster having a larger number density of these `bright' sources.

Work by \citet{Anella13} has been done to show that if a new powerful timing pulsar was discovered in the direction of the Virgo cluster, a PTA would have twice the sensitivity to a region about $20^o$ around the discovered pulsar. Given that the sensitivity of a single pulsar to a GW source falls off as $1 + cos \theta$, where $\theta$ is the angle between a pulsar and a GW source \citep{Burt10}, we recommend focusing the search for new pulsars in the vicinity of the Virgo cluster.

As stated earlier in this paper, this is only a broad estimate that assumed an equal probability for all galaxies to contain a binary with equal mass black holes. 
Future work is planned to incorporate work done by \citet{Rosado13}, which uses the Sloan Digital Sky Survey and the Millenium simulation data \citep{Springel05} to search for SMBHBs in the redshift range of $z = 0.01 - 0.7$. Combining these two data sets is the next step towards creating realistic population distributions for single source   GWs detectable by PTAs.
While the distribution of detectable single sources will most likely scale with any new estimate, $\langle P \rangle$ is more affected by these specific probabilities, and a future paper will address these factors.

\acknowledgements{This work was supported by the NSF Partnerships in International Research and Education (PIRE) Grant No. 0968296 (http://nanograv-pire.wvu.edu/) and by NSF CAREER Award 07-48580 to A. Lommen.  This research has made use of MEALpix developed by GWAstro Research Group and available from http://gwastro.org. We acknowledge the usage of the HyperLeda database (http://leda.univ-lyon1.fr)}

\pagebreak

\appendix
\onecolumngrid

In this section, we will derive two statistics which can be calculated from existing survey data and are proportional to two measures of gravitational wave strength in a given direction. The two measures are the number of detectable binary systems and the strength of the stochastic GW signal coming from a particular region of the sky. We will denote these quantities as $dN/d\Omega$ and $d h^2/df d\Omega$, respectively. Our calculation starts with the differential rate of SMBH coalescence given by $R = dN/dtdM_Cdzd\Omega$. This quantity represents the number of SMBH binary systems coalescing per unit observer time, per unit chirp mass, $M_c$, of the system, per unit red shift z, occurring within a solid angle $d\Omega$ as seen by an observer at Earth. Given the amount of time, $d\tau$, that a binary system spends emitting a GW with frequency between $f$ and $f+df$, the number of detectable binary systems per unit solid angle may be expressed as:
\begin{equation}
\label{dndomega}
\frac{d N}{d\Omega} = \int R \frac{d\tau}{df} dz dM_c df.
\end{equation}
In the above, the integration should be performed over that region of $z$, $M_c$, and $f$ where GWs would be detectable by a given PTA configuration.

Given the amplitude of the gravitational wave strain, $h_s$, emitted by a particular SMBH system, the strength of the stochastic GW signal may be written as:
\begin{equation}
\label{dh2domega}
\frac{d h^2}{df d\Omega} = \int R \frac{d\tau}{df} h_s(f,z,M_c)^2 dz dM_c. 
\end{equation}
We want to make estimates of the relative strength of both $dN/d\Omega$ and $d h^2/d\Omega$ using observational data from galaxy surveys. In order to do this, we will make the following assumptions:  1) $R$ is proportional to the number of observable galaxies per unit solid angle in a given direction, 2) the evolution of the SMBH binaries are dominated by the effects of GW emission, 
3) the probability of a galaxy harboring a SMBH is the same for all galaxies, 4) the chirp mass of the binary system is proportional to the total luminosity of the galaxy, $L_t$, 5) all galaxies of interest have $z<<1$. 

Using the fact that $R$ is independent of frequency,  equation \ref{dndomega} may be written as:
\begin{equation}
\label{dndomega2}
\frac{d N}{d\Omega} = \int R \Delta \tau(M_c,z) dz dM_c,
\end{equation}
where $\tau(M_c)$ is the total time a SMBH with chirp mass $M_c$ is detectable in a given PTA and is given by:
\begin{equation}
\tau(M_c,z) = \frac{5}{256}\left(\frac{c^3}{G M_c}\right)^{5/3} \pi^{-8/3} \left(f_l^{-8/3} - f_h^{-8/3}\right).
\end{equation}
The frequencies $f_l$ and $f_h$ are the lowest and highest detectable frequencies given the sensitivity of a PTA and a particular SMBH binary system. Note that both $f_l$ and $f_h$ depend on $M_c$ and $z$. These frequencies are calculated using the following expression for the GW strain amplitude:
\begin{equation}
\label{hs}
h_s(f,z,M_c) = 4 \sqrt{\frac{2}{5}} \frac{(G M_c)^{5/3}}{c^4 D(z)} f^{2/3} (1+z)^{2/3} \pi^{2/3},
\end{equation}
together with an expression for the minimum detectable strain of a PTA. 

Let $R_g = dN_g/dL_t dz d\Omega$ be the total number of galaxies per unit total luminosity, per unit red shift, per unit solid angle. Assumptions one and three allow us to write $R = C_1 R_g$ where $C_1$ is a constant. The total number of galaxies per unit solid angle in a given direction is given by
\begin{equation}
\frac{dN_g}{d\Omega} = \int R_g dL_t dz.
\end{equation}
With assumption 4, we can write $M_c = C_2 L_t$ where $C_2$ is a constant. Putting this all into equation \ref{dndomega2}, we have
\begin{equation}
\frac{dN}{d\Omega} = C_1 C_2 \int R_g \Delta \tau(C_2L_t,z) dz dL_t.
\end{equation}
Given that $R_g/ dN_g/d\Omega dL_t dz$ is the fractional number of galaxies in a given direction with total luminosity between $L_t$ and $L_t + dL_t$ located at a red shift between $z$ and $z+dz$, we can write the detectable number of systems as
\begin{equation}
\frac{dN}{d\Omega} = C_1 C_2 \frac{dN_g}{d\Omega} <\tau(C_2 L_t,z)>
\end{equation}
where the angle brackets represent an average over all galaxies in a particular direction. Since $\tau(M_c,z) \propto M_c^{-5/3} (f_l^{-8/3} - f_h^{-8/3})$, we find that
\begin{equation}
\frac{dN}{d\Omega} \propto F^\prime = \frac{dN_g}{d\Omega} <L_t^{-5/3} \left(f_l^{-8/3} - f_h^{-8/3}\right)>,
\end{equation}
where $F^\prime$, defined in the above equation, is a quantity that may be estimated from galaxy catalogue data and is directly proportional to the number of detectable sources per unit solid angle in a given direction. We can also define an idealized statistic which is valid for the case where the PTA can detect all frequencies down to some cuttoff frequency. In this case, we can ignore the frequency dependent terms and use 
\begin{equation}
F = \frac{dN_g}{d\Omega} <L_t^{-5/3}>.
\end{equation}
Like $F^\prime$, $F$ is proportional to $\frac{dN}{d\Omega}$ but it only depends on the properties of the galaxy distribution and not the details of a particular PTA. An estimate for $F$ is simply the sum of $L_t^{-5/3}$ over all galaxies in a particular small region of the sky.

Using the same assumptions as with equation \ref{dh2domega}, one can show that:
\begin{equation}
\frac{d h^2}{df d\Omega} = C_1 C_2 \int R_g \frac{d\tau}{d f} h_s(f,z,C_2 L_t)^2 dz dL_t df.
\end{equation}
Here, the range of integration is over all $z, L_t$ at a fixed frequency $f$, unlike the expression for $dN/d\Omega$. Using assumption 2), we know that  
\begin{equation}
\frac{d\tau}{df} = \frac{5}{96}\left(\frac{c^3}{G M_c}\right)^{5/3} \pi^{-8/3} f^{-11/3}\left(1 + z\right)^{-5/3}.
\end{equation}
This, together with equation \ref{hs} gives us the following:
\begin{equation}
\frac{d h^2}{df d\Omega} \propto P = \frac{d N}{d\Omega}<\frac{L_t^{5/3}}{D(z)^2}>,
\end{equation}
where $P$ as defined above may be estimated from existing data and is proportional to the strength of the stochastic GW emission in a particular direction. An estimate for $P$ is obtained by summing $L_t^{5/3}/D(z)^2$ over all galaxies in a particular direction.
 
Maps of both the number of detectable sources, $F$ and the strength of the stochastic GW signal, $P$, calculated using data from extragalactic databases are presented in \S 4.


\end{document}